\documentclass[11pt]{article}
\usepackage{latexsym}

\def\be{\begin{equation}}
\def\ee{\end{equation}}
\def\bea{\begin{eqnarray}}
\def\eea{\end{eqnarray}}

\def\case#1/#2{\textstyle\frac{#1}{#2}}

\begin{document}
\begin{titlepage}

\vspace{.7in}

\begin{center}
\Large\bf {DYNAMICS OF PURE SHAPE, RELATIVITY AND THE PROBLEM OF TIME}\normalfont
\\
\vspace{.4in} \normalsize \large{Julian
Barbour$^1$}
\\
\normalsize \vspace{.4in}

$^1$ College Farm, South Newington, Banbury, Oxon, OX15 4JG, UK

\end{center}
\vspace{.4in} 

\begin{abstract}

A new approach to the dynamics of the universe based on work by \'{O} Murchadha, Foster, Anderson and the author is presented. The only kinematics presupposed is the spatial geometry needed to define configuration spaces in purely relational terms. A new formulation of the relativity principle based on Poincar\'e's analysis of the problem of absolute and relative motion (Mach's principle) is given. The entire dynamics is based on shape and nothing else. It leads to much stronger predictions than standard Newtonian theory. For the dynamics of Riemannian 3-geometries on which matter fields also evolve, implementation of the new relativity principle establishes unexpected links between special relativity, general relativity and the gauge principle. They all emerge together as a self-consistent complex from a unified and completely relational approach to dynamics. A connection between time and scale invariance is established. In particular, the representation of general relativity as evolution of the shape of space leads to a unique dynamical definition of simultaneity. This opens up the prospect of a solution of the problem of time in quantum gravity on the basis of a fundamental dynamical principle.

\end{abstract}

\vspace{.3in} 

Electronic address:{$^1$ julian@platonia.com} 

\end{titlepage}

\section{Introduction}

In this paper, I wish to discuss the foundations of cosmology and our notion of time. The goal is to contribute to the creation of a quantum theory of the universe. I shall draw attention to some conceptual issues that in my opinion have not hitherto been adequately discussed. Since this is a contribution to an interdisciplinary workshop, I shall keep the discussion as simple as possible. In fact, the model that I shall present, which treats non-relativistic particles in Euclidean space, hardly seems realistic. However, it contains the simplest implementation of a dynamical variational principle that can also be applied in general relativity, as I shall outline in more qualitative terms. It is the principle that I wish to explain. Moreover, the problems to be addressed are so challenging, there is much to be said for attacking them initially in the simplest possible situations. I also happen to believe that anyone competent in quantum mechanics should be interested in the issues raised by the model independently of the quantum cosmological significance. The question is this: Can one do meaningful quantum mechanics with significantly less external kinematic structure than is currently employed in quantum theory? This is a topic ideally suited to this interdisciplinary volume.

Some historical background will be helpful. In order to formulate his laws of motion, Newton introduced a rigid external framework: absolute space and time. This framework was simply taken over in its entirety by the creators of quantum mechanics. It was only somewhat modified to accommodate quantum field theory, and a fixed framework is still deeply embedded in that theory. However, ten years before the creation of quantum mechanics, Einstein had created the general theory of relativity. There is universal agreement that Einstein to a very large degree abolished Newton's absolute framework. There is less agreement about what, if anything, replaced it in the classical theory and what kind of framework is appropriate for a quantum theory of gravity. This, I believe, is the main reason why people are still seeking the foundations of quantum cosmology. I want to argue for a new and clearer formulation of the relativity principle. This will lead me to the notion of \textit{dynamics of pure shape}. 

It will be helpful to distinguish two different notions of relativity. There is, first, the intuitive idea that position can only be defined relative to observable objects. Suppose we were to contemplate a universe consisting of $N$ point particles in Euclidean space with separations $r_{ij}$ between them. The opponents of Newton's absolute space and time, above all Leibniz \cite{Leibniz} and Mach \cite{Mach}, argued that dynamics should be directly formulated in terms of the $r_{ij}$. Let me call this \textit{kinematic relativity}. In 1902, Poincar\'{e}  \cite{Poincare1} pointed out that in fact Newtonian point-particle dynamics can always be formulated in terms of the $r_{ij}$, but that then the structure of its initial-value problem is changed compared with the formulation in absolute space (or, in modern terms, an inertial system). This is such an important issue that it needs to be spelled out. If we are given initial positions $\textbf{x}_i$ and initial velocities $\dot\textbf{x}_i$ in an inertial system (together with the masses $m_{i}$ and the force law), then Newton's laws determine the past and future motions uniquely. The situation is characteristically different if one is given $r_{ij}$ and $\dot r_{ij}$. The fact is that such data contain no information at all about the overall rotation of the system. One cannot determine the angular momentum $\textbf M$ of the system, and for different $\textbf M$ very different evolutions result. Initial data that are identical from the point of view of kinematic relativity give rise to different evolutions. Moreover the defect has an odd structure. Suppose $N$ is large. In three-dimensional Euclidean space, the $r_{ij}$ contain $3N-6$ independent data, and their time derivatives contain the same number. This number is the order of a million for a globular cluster. Just three more numbers are needed if the evolution is to be uniquely determined. They could be three of the second time derivatives. But why does one need three and not $3n-6$? Poincar\'{e} argued, persuasively in my view, that the only valid objection to Newton's use of absolute space and time resided in this curious need to specify a small number of extra data. He said that if one could formulate a relational dynamics (i.e., one containing only $r_{ij}$ and its time derivatives) free of this defect, the problem of absolute motion would be solved. Poincar\'{e} made no attempt to find such a dynamics. Moreover, he pointed out that the empirical evidence seemed to show conclusively that nature did not work in this way.

I shall argue that Poincar\'{e}'s analysis is very sound and that empirically adequate theories satisfying a criterion along the lines he proposed can be formulated. However, it will be helpful to push his analysis somewhat further, which I shall do in the next section. For the moment, let me merely note that if one takes kinematic relativity seriously one will wish to formulate a dynamics whose initial-value problem satisfies a well-defined criterion. For reasons that will soon become apparent, let me call this the \textit{constructive approach} or \textit{Poincar\'{e}'s relativity principle}. It can be contrasted with the approach that was adopted by Einstein in creating both special and general relativity and is known as the \textit{principle approach}. It grew out of a generalization of \textit{Galilean relativity}, which Einstein transformed into the restricted relativity principle. In order to construct a theory in which uniform motion could not be detected, Einstein made no attempt to formulate a complete theory of particles interacting with Maxwell's electromagnetic field. Instead, he postulated the existence of a family of distinguished (inertial) frames all in uniform translational motion relative to each other and required the laws of nature to take the identical form in all of them. The dramatic results that he obtained came from combination of this relativity principle with his postulate about the behavior of light. As he explained in his Autobiographical Notes \cite{AutoNotes} (see also \cite{Barbour}, on which this discussion is based), he was encouraged to adopt this approach because of the success of phenomenological thermodynamics based based on `impotence' principles: the impossibility of constructing perpetual motion machines of the first and second kind. The impotence in his case was the impossibility of detecting uniform motion through the ether (or absolute space) by means of processes that unfold in a closed system.

In 1907, Einstein realized that the equivalence principle enabled him to extend the restricted relativity principle to include another `impotence' -- the inability to detect uniform acceleration. This insight was decisive. Some years earlier, Einstein had read Mach's critique of Newton's absolute space and time and was extremely keen to reformulate dynamics along the broad lines advocated by Mach. He wanted to show that absolute space did not correspond to anything in reality -- that it could not be revealed by any experiment. Special relativity had shown that uniform motion -- relative to the ether or to absolute space -- could not be detected by any physical process. The equivalence principle suggested to him that it might be possible to extend his relativity principle further. If he could extend it so far as to show that the laws of nature could be expressed in identical form in all conceivable frames of reference, this requirement of general covariance would ``[take] away from space and time the last remnant of physical objectivity''\cite{Einstein16}. He would have achieved his Machian aim. Unfortunately, within two years Einstein had been forced by a critique of Kretschmann \cite{Kretschmann} to acknowledge that any physical theory must, if it is to have any content, be expressible in generally covariant form. He argued \cite{Einstein18} that the principle nevertheless had great heuristic value. One should seek only those theories that are \textit{simple} when expressed in generally covariant form. However, Einstein gave no definition of simplicity. Since then, and especially as a result of quantization attempts, there has been a vast amount of inconclusive discussion about the significance of general covariance \cite{BarPlanck} and its implications for quantization. A point worth noting is that Einstein treated space and time as a single unit and considered general coordinate transformations on a four-dimensional manifold. The distinction between space and time and the manner in which they are to be treated is very largely erased.

In several recent papers \cite{BOF, AB, SIG, ABFO}, my collaborators and I have developed what we call \textit{the 3-space approach}. It is based on a generalization of Poincar\'{e}'s relativity principle and casts new light on this issue. Above all, it replaces Einstein's vague simplicity requirement by a well-defined constructive principle based on the amount of data needed to formulate initial-value problems. In addition, space and time are treated in completely different ways. This might seem to be a retrogressive step, but actually the approach not only achieves everything that Einstein did by presupposing a four-dimensional unity of space and time but even more. In a real sense it explains why there is a universal light cone (which Einstein presupposed) and why the gauge principle holds. The details of this work goes beyond the scope of the present paper, though I will indicate how these results are obtained in the final section. Readers wishing for full details are referred to the original papers. In this paper, I merely wish to get across the basic ideas and draw attention to some of the interesting possibilities that arise.

\section{Basic Ideas}

The 3-space approach is a natural modification of the basic scheme employed in the variational principles of mechanics \cite{Lanczos}. The main difference is that the 3-space approach uses less kinematic structure. Let us first consider the notion of configuration space. In the Newtonian $N$-body problem, this is defined relative to an inertial system, which serves two purposes. First, it provides a definition of \textit{equilocality} at different instants of time: relative to the system, one can say that a particle is at the same position at two different times. Displacements are then well defined. This is essentially the reason why Newton introduced absolute space. Second, an inertial system brings with it a notion of time difference. Given a notion of simultaneity, so that one can say that the system has some instantaneous configuration at a given instant, that still does not say `how much time there is' between two different configurations in a history. This extra information is supplied with an inertial system.

Let us now see how much of this structure can be shed without making dynamics impossible. Suppose a universe consisting of $N$ point particles in Euclidean space. It is rather natural to assume that only the relative configuration counts as physical reality. One would like to say that all configurations that can be carried into exact congruence by Euclidean translations and rotations are the same physical configuration. This amounts to quotienting the Newtonian configuration space Q of $3N$ dimensions by these six symmetries (three translations, three rotations) to obtain the \textit{relative configuration space} (RCS) Q$_{\scriptsize\textrm{RCS}}$ of $3N-6$ dimensions \cite{BB}. It is also very natural to go one step further and say that size is relative. Then two configurations that can be made congruent by the action of translations, rotations and dilatations are to be regarded as identical. The corresponding quotient space has $3N-7$ dimensions and may be called \textit{shape space}: Q$_{\scriptsize\textrm{SS}}$.

A Newtonian dynamical history is a curve in Q traversed at a certain speed relative to some external measure of time. Now time must always be deduced from the motion of some object in the universe that is itself subject to the laws of nature. Therefore, if the system that we are considering is the entire universe, the information from which we deduce time must in fact already be encoded in the curve in Q. We shall see how this is done shortly. But this observation itself already suggests that, if we are considering the dynamics of the entire universe, it ought to be sufficient to set up a theory of \textit{curves} in the configuration space and dispense with the idea that they are traversed at some speed. The simplest curves are those whose determining law is such that an initial point and initial direction suffice to determine the entire curve. They are geodesics. The key difference from Newtonian curves is that only the initial direction, not the initial direction and the speed in that direction, needs to be specified. This is a natural extension to Poincar\'{e}'s relativity principle. We seek laws that determine histories with the minimum number of initial data. Such laws are \textit{maximally predictive}. 

Going from curves traversed at speed to geodesics is one way of the two ways in which we can reduce the number of initial data. The second is to formulate Poincar\'{e}'s relativity principle on the smaller quotient spaces Q$_{\scriptsize\textrm{RCS}}$ and Q$_{\scriptsize\textrm{SS}}$. One of the main points I want to make in this paper is that there is a surprising interconnection between these two ways of reducing initial data. The first may be called the elimination of time and the second elimination of potentially redundant geometrical structure. We shall see that the elimination of time is not truly effective unless the employed geometrical structure is pared down to the absolute minimum. There is an unexpected connection between time and geometry, specifically scale invariance.

An overall characterization of the 3-space approach is here appropriate. The basic idea is to postulate the geometrical structure of space, which is assumed here to have three dimensions but in principle any dimension is possible. This paper will mainly be concerned with Euclidean space, though much more interesting possibilities arise if space is Riemannian. Because of length restrictions, I shall only be able to discuss them briefly and will concentrate on the Euclidean case in order to get the idea across. We assume that the space we consider is occupied by geometrical objects, which may either be point particles or fields. These objects and the space in which they reside define configurations and associated configuration spaces. The key point, as already explained, is that one obtains a hierarchy of configuration spaces by quotienting with respect to the symmetries inherent in the space that is presupposed. The mathematical existence of these quotient spaces seems to reflect a deep property of the world. They seem to be a necessary concomitant of the existence of spatial order. The results so far obtained in the 3-space approach suggest that the structure of the quotient spaces determines much more of the fundamental laws of classical physics than has hitherto been supposed. The point is that it is not easy to construct geodesic principles on quotient spaces. If they are to be consistent, they impose strong restrictions.

There exists more than one way to construct geodesic principles on quotient spaces, but one of them seems to be clearly distinguished compared with all the others on account of its geometrical nature. This is based on a principle that we call \textit{best matching}. If we are to define geodesics on any space, we need to define a metric on it. We need to define a distance between any two pairs of neighboring points of the space. The points in our case are complete configurations of the system that we are considering. They necessarily have some small intrinsic difference. Now the basis of geometry is congruence. However, two intrinsically different configurations cannot be brought to exact congruence. Best matching of two such configurations is based on the idea of bringing them, in a well-defined sense, as close as possible to exact congruence and using the `mismatch' from it to define the distance between them. Geodesics can then be defined with respect to this best-matching metric. As we shall see, the consistent application of this idea leads to interesting restrictions. Before proceeding with the formal development, let me give an idea of the nature of these restrictions.

The benchmark for considering them is standard Newtonian theory. Suppose we take a generic solution of the Newtonian $N$-body problem of celestial mechanics and successively `throw away information'. First, instead of giving the time at which the successive configurations are realized we can simply give their sequence. Next, we can omit the information, characterized by six numbers, that specify the overall position and orientation of the system. We can do this by giving only the inter-particle separations $r_{ij}$. Finally, we can omit the scale information (one number) contained in the $r_{ij}$. This is most conveniently done by normalizing them by the square root of the moment of inertia $I$ about the center of mass: 
\begin{equation}
I=\sum_i m_i \textbf{x}_i^2={1\over M}\sum_{i<j}m_im_jr_{ij}^2, M=\sum_i m_i.
\label{MOI}
\end{equation}

The resulting information can be plotted as a curve in shape space. If this curve were a geodesic, it would be determined by specification of an initial point and initial direction in shape space. For the 3-body problem, shape space has two dimensions, so in this case one would need three numbers: two to specify the initial point, one to specify the initial direction. This is the ideal that must be met by a dynamics of pure shape. It turns out that a generic Newtonian solution needs no less than five further numbers to be fully specified. It is illuminating to consider what they are. 

First, it should be noted that in shape space all dimensional information is lost. We have no knowledge of length scales or clock rates. Now the fundamental dynamical quantities in Newtonian theory such as energy, momentum, and angular momentum, as well as Newton's gravitational constant G, depend on these scales, but only scale-invariant quantities can affect the form of the Newtonian curves projected down to shape space. Let us consider what they are. First, at any instant the angular momentum vector has a certain direction relative to the instantaneous configuration. Two numbers are associated with this information. Next, the Newtonian kinetic energy can be decomposed into a part associated with overall rotation, a part associated with change of shape, and a part associated with change of size. Two independent scale-invariant ratios can be formed from them. These are four of the five numbers. The final number is in many ways the most enigmatic. It is the instantaneous ratio $H=T/V$ of the kinetic energy $T$ to the potential energy $V$. Intuitively it exists in Newtonian theory because the external time makes it possible to convert displacements into velocities.

We shall see that the transition from Q to Q$_{\scriptsize\textrm{RCS}}$ ensures that the rotational motion associated with angular momentum no longer plays a role. This eliminates three of the above five numbers. However, two still remain. Rather remarkably, both are eliminated when we take the further step to a dynamics of pure shape on Q$_{\scriptsize\textrm{SS}}$. One of them measures the kinetic energy associated with change of size, and it is no surprise that this is eliminated in a dynamics of pure shape. But the other, related to the energy, seems intuitively to have something to do with time. After all, time and energy form a canonical pair in Hamiltonian dynamics \cite{Lanczos}. It is therefore surprising that a scaling requirement appears to have a bearing on time. I shall come back to this later.

In the next section, I shall discuss the formulation of geodesic principles in a way that highlights the difference from Newtonian theory. In Sec. 4, I shall explain the technique of best matching and in Sec. 5 show how Newtonian theory can be recovered to excellent accuracy from a scale-invariant theory. In Sec. 6, I shall indicate how these ideas can be applied to Riemannian geometry and fields and yield a new perspective on general relativity.

\section{Jacobi's Principle}
The first step to a dynamics of pure shape is the elimination of
time by Jacobi's principle \cite{Lanczos}, which describes all 
Newtonian motions of one value $E$ of the total energy as
geodesics on configuration space. Further discussion of the implications of Jacobi's principle can be found in \cite{BOF, CQG94, EOT}.  

For $N$ particles of masses 
$m_{i}$ with potential
$U(\textbf{x}_{1}, \dots , \textbf{x}_{N})$ and energy $E$, the
Jacobi action is \cite{Lanczos}
\begin{equation}
    I_{\scriptsize\textrm{Jacobi}} =    2\int\sqrt{E -
    U}\sqrt{\tilde T} \textrm{d}\lambda,
\label{Jacobi}
\end{equation}
where $\lambda$ labels the points on trial curves and $ \tilde{T}
= \sum {m_{i} \over
2}{\textrm{d}{\textbf{x}}_{i}\over\textrm{d}\lambda}\cdot
{\textrm{d}{\textbf{x}}_{i}\over\textrm{d}\lambda}$ is the
parametrized kinetic energy. The action (\ref{Jacobi}) is timeless
since the label $\lambda$ could be omitted and the mere
displacements $d\textbf{x}_{i}$ employed, as is reflected in the
invariance of $I_\textrm{\scriptsize{Jacobi}}$ under the
reparametrization
\begin{equation}
    \lambda \rightarrow f(\lambda).\label{rep}
\end{equation}

In fact, it is much more illuminating to write the Jacobi action in the form
\begin{equation}
    I_{\scriptsize\textrm{Jacobi}} =    2\int\sqrt{E -
    U}\sqrt{T^*},\hspace{.5cm}T^*=\sum {m_{i} \over
2}\textrm{d}{\textbf{x}}_{i}\cdot
{\textrm{d}{\textbf{x}}_{i}},
\label{Jacobi*}
\end{equation}
which makes its timeless nature obvious and dispenses with the label $\lambda$.

The characteristic square roots of $I_{\scriptsize\textrm{Jacobi}}$
fix the structure of the canonical momenta:
\begin{equation}
    \textbf{p}_{i} =
    {\partial {\cal L} \over \partial(\textrm{d}\textbf{x}_{i}/\textrm{d}\lambda)} =
    m_{i}\sqrt{E - U \over {\tilde T}} {d \textbf{x}_{i}
    \over d \lambda},
\label{CanMom}
\end{equation}
which, being homogeneous of degree zero in the velocities, satisfy
the constraint \cite{Dirac}
\begin{equation}
    \sum{{{\textbf{p}}_{i}}\cdot
    {{\textbf{p}}_{i}}\over 2m_{i}}-E+U=0.
\label{QuadCon}
\end{equation}

The Euler--Lagrange equations are
\begin{equation}
    {\textrm{d}\textbf{p}^{i} \over \textrm{d} \lambda}= {\partial
    {\cal L} \over \partial \textbf{x}_{i}} = -\sqrt{{\tilde T} \over
    E - U}{\partial U\over
    \partial \textbf{x}_{i}} ,
\label{JacobiEL}
\end{equation}
where $\lambda$ is still arbitrary. If we choose it such that
\begin{equation}
    {{\tilde T} \over E - U} = 1 \Rightarrow {\tilde T} = E - U
\label{EnCon}
\end{equation}
then (\ref{CanMom}) and (\ref{JacobiEL}) become
$$
    \textbf{p}_{i} = m_{i}{{\textrm{d}\textbf{x}_{i}} \over \textrm{d}
    \lambda},\hspace{1.0cm} {\textrm{d} \textbf{p}_{i} \over
    \textrm{d}\lambda} = -{\partial U\over\partial \textbf{x}_{i}},
$$
and we recover Newton's second law w.r.t this special $\lambda$.
However, (\ref{EnCon}), which is usually taken to express energy
conservation, becomes the \textit{definition of time}. Indeed,
this emergent time, chosen to make the equations of motion take
their simplest form \cite{Poincare2}, is the astronomers'
operational ephemeris time \cite{Clemence}. It is helpful to see how `change creates
time'. The increment $\delta t$ generated by displacements
$\delta \textbf{x}_{i}$ is
\begin{equation}
    \delta t={\sqrt{\sum
    m_{i}\delta\textbf{x}_{i}\cdot\delta\textbf{x}_{i}}\over\sqrt{2(E-U)}}
    \equiv{\delta s\over\sqrt{2(E-U)}}.
\label{clem}
\end{equation}
Each particle `advances time' in
proportion to the square root of its mass and to its displacement, the total
contribution $\delta s$ being weighted by $\sqrt{2(E-U)}$. 

In the previous section, I discussed the role of the energy in determining the curves of generic Newtonian solutions when projected down to shape space. The Jacobi action (\ref{Jacobi*}) illuminates this issue. Considered purely mathematically, $T^*$ by itself already defines a (Riemannian) metric on Q. It is the kinetic metric \cite{Lanczos}. The function $(E-U)$ multiplying $T^*$ is a conformal factor that transforms the original kinetic metric, which describes pure inertial motion, into a conformally related metric. It is this conformal factor that introduces forces and the effect of the energy into Newtonian mechanics. The decomposition of the conformal factor into the constant $E$ and the conventional Newtonian potential $-U$, which is a function of the inter-particle separations, is artificial from this point of view. In the development of best matching in the next section, it will be best to start by allowing the conformal factor to be an arbitrary function on Q.

\section{Best Matching}

The idea of best matching is simple and arises from a very natural problem: How can one quantify the difference between two nearly identical configurations in an intrinsic manner? No additional structure like an inertial system is to be used. In addition, a universally applicable method is required. This is explained in detail in \cite{SIG, ABFO}. Here I will explain the gist of the method for the case of the $N$-body problem. Represent the two configurations in the same coordinate grid. In configuration 1, particle $i$ will have coordinates $\textbf{x}_i$. In configuration 2, it will have coordinates $\textbf{x}_i+\textrm d\textbf{x}_i$. Now consider the quadratic form
\begin{equation}
F\sum_i m_i \textrm d\textbf{x}_i\cdot\textrm d\textbf{x}_i,
\label{a}
\end{equation}
where the conformal factor $F$, assumed positive since we are going to take the square root of (\ref a) in order to obtain a Jacobi-type action, can in principle be an arbitrary function of the coordinates $\textbf x_i$.\footnote{Clearly, a more general, non-diagonal form could be assumed instead of (\ref a). This is one of several issues within the 3-space approach that are currently being studied \cite{Variations}.} We can now use the Euclidean generators of translations, rotations and dilatations separately on each of the configurations, generating different `placings' of them, and calculate (\ref{a}) for each change made to the pair of configurations. The points they define in shape space will be unchanged by these operations, which merely affect their mathematical representation. The idea of best matching is to seek the minimum of (\ref a) with respect to all possible placings and to declare this to be the metric distance between the configurations. If a consistent scheme is to be obtained, interesting restrictions arise. It is easier to visualize best matching in the finite-difference form just described. However, calculations are more readily done with continuous variations, which correspond to a Jacobi action of the form
\begin{equation}
I_{\scriptsize\textrm{BM}}=2\int\textrm d\lambda\sqrt F\sqrt T, \hspace{.5cm}T=\sqrt{{1\over 2}\sum_i m_i\left ({\textrm d\textbf x_i\over \textrm d\lambda}-\textbf c_i\right )\cdot \left ({\textrm d\textbf x_i\over \textrm d\lambda}-\textbf c_i\right )},
\label{b}
\end{equation}where $\textbf c_i$, which has the dimensions of a velocity, is the correction that arises from $\lambda$-dependent transformations on the instantaneous configuration generated by translations, rotations and dilatations. For example, consider a $\lambda$-dependent translation $\textbf x_i\rightarrow \textbf x_i+\textbf b(\lambda).$ It generates the velocity transformation $\dot\textbf x_i\rightarrow \dot\textbf x_i+\dot\textbf b(\lambda)$, where $\dot\textbf x_i=\textrm d\textbf x_i/\textrm d\lambda$. If we make such transformations, (\ref b) without the correction terms will be changed in an arbitrary manner by the arbitrary vector function $\dot\textbf b$. To counteract this effect of translations, we take the correction $\textbf c_i$ to be an arbitrary vector function $\textbf a$ and vary the action (\ref b) with respect to it as a Lagrange multiplier. To counteract the effect of simultaneous arbitrary translations, rotations and dilatations, we take the correction to be
\begin{equation}
\textbf c_i=\textbf a+\omega \times\textbf x_i+D\textbf x_i
\label{c}
\end{equation}and vary with respect to the vector functions $\textbf a$ and $\omega$ and the scalar function $D$ as Lagrange multipliers. This variation leads to constraints satisfied by the canonical momenta $\textbf p_i$,
\begin{equation}
\textbf p_i=\sqrt{F\over T}m_i\left ({\textrm d\textbf x_i\over \textrm d\lambda}-\textbf c_i\right ),
\label{d}
\end{equation}
of the physical variables $\textbf x_i$. The constraints that arise from the translations, rotations and dilatations are, respectively, 
\begin{equation}
\textbf P\equiv\sum_i\textbf p_i=0,
\label{e}
\end{equation}
\begin{equation}
\textbf M\equiv\sum_i\textbf x_i\times\textbf p_i=0,
\label{f}
\end{equation}
\begin{equation}
v\equiv\sum_i\textbf x_i\cdot\textbf p_i.
\label{g}
\end{equation}

Now comes a crucial point. Do the Euler--Lagrange equations,
\begin{equation}
{\textrm d\textbf p_i\over\textrm d\lambda}=\sqrt{T\over F}{\partial F\over\partial \textbf x_i},
\label{h}
\end{equation}propagate the constraints? The simple calculation shows that (\ref e) will propagate only if $F$ is translationally invariant, (\ref f) will propagate only if $F$ is rotationally invariant, and (\ref g) will propagate only if $F$ is homogeneous of degree -2.\footnote{These conditions, derived here as consistency requirements, are necessary consequences of the fact that best matching is being used to define a metric on a quotient space. In fact, as is shown in \cite{SIG}, each symmetry with respect to which best matching is performed leads to two conditions: a linear constraint on canonical momenta and a condition on the potential $F$ that ensures its propagation.}
Note that, as is described in detail in \cite{SIG}, the linear constraints (\ref e), (\ref f), and (\ref g) owe their existence to the fact that (\ref b) is invariant under $\lambda$-dependent translations, rotations, and dilatations provided one defines the transformation law of the three correction terms in $\textbf c_i$ to be the same as that of the velocities but with the opposite sign. Besides the three linear constraints, there is also a quadratic constraint analogous to (\ref{QuadCon}):
\begin{equation}
    \sum{{{\textbf{p}}_{i}}\cdot
    {{\textbf{p}}_{i}}\over 2m_{i}}-F=0.
\label{i}
\end{equation}

This model, with linear constraints that are uniquely determined by the symmetries of space and a quadratic constraint that follows directly from the idea that time is redundant if the dynamics of the universe (as opposed to subsystems of it) is considered, is interesting from several points of view. Before we discuss them, some preparatory remarks are in order. First, the constraints apply only to the complete system of particles treated as an `island universe'. Subsystems are not constrained. It is only necessary that the contributions of all subsystems sum to zero. Second, it is always possible to employ a coordinate frame in which the corrections terms $c_i$ vanish and a special `time' label $\lambda$ for which $F/T=const$. Then the Euler--Lagrange equations are identical to Newton's equations in an inertial system. Third, the first two linear constraints tell us that in this preferred system the Newtonian momentum and angular momentum of the universe vanish. Because of the Galilean invariance of Newtonian mechanics, the vanishing of the momentum is not a new physical prediction. It is however derived within the logic of best matching. The vanishing of the angular momentum is not enforced by Newton's equations, the rotational symmetry of which only ensures conservation of angular momentum. Best matching enforces both the symmetry and the exact vanishing of the conserved quantity.

Now we come to consider the third linear constraint. This is by far the most drastic in its consequences and also impacts on the quadratic constraint (\ref i). It introduces a new conserved quantity in Newtonian dynamics, which however, from the point of view of Newtonian dynamics, occurs only under very special circumstances. It has long been recognized by $N$-body specialists that potentials homogenous of degree -2 in the inter-particle separations represent an interesting special case. This follows from the so-called Lagrange--Jacobi relation \cite{SIG}, which gives an universal expression for the time variation of the moment of inertia (\ref{MOI}) in any case in which the the potential is homogenous of degree k:
\begin{equation}
    \ddot{I}=4(E-U)-2kU.
\label{Iddh}
\end{equation}
Consider Newtonian celestial
mechanics, for which $k=-1$. Then $\ddot{I}=4E-2U$, from which Lagrange deduced the
first qualitative result in dynamics. Since $U<0$ for gravity, $E\geq 0$ implies $ \ddot I>0.$ Thus $I$ is concave
upwards and must tend to infinity as $t\longrightarrow +\infty$
and $t\longrightarrow -\infty$. In turn, this means that at least
one of the interparticle distances must increase unboundedly, so
that any system with $E\geq 0$ is unstable.

Another consequence of (\ref{Iddh}) is the virial theorem. For
suppose that the system has virialized, so that $I\approx0$. Then
$4E=(2k+4)U.$

For our purposes, the most interesting consequence of (\ref{Iddh})
arises when $k=-2$. For then
\begin{equation}
    \ddot I=4E.
\label{Ih-2}
\end{equation}

Thus, $I$ has the parabolic dependence $I=2Et^{2}+bt+c$ on the
time and will tend rapidly to zero or infinity. Such a system is
extremely unstable, either imploding or exploding.

However, suppose $E=0$. Then $\ddot I=0$ by (\ref{Ih-2}), so that
\begin{equation}
\dot I=2\sum m_{i}\dot\textbf x_{i}\cdot \textbf x_{i}=2\sum
\textbf p_{i}\cdot \textbf x_{i}=\textrm{constant}.
\label{j}
\end{equation}
Thus, $v=\sum \textbf p_{i}\cdot \textbf x_{i}$ is a \textit{new conserved quantity}. I am not aware that it has been given any definite name in the literature (or even that its potential significance in dynamics has been recognized). Since it has the same dimensions (action) as angular momentum and is closely analogous to it, I have called it in \cite{SIG} the \textit{expansive momentum}. It is precisely the quantity that we have found must vanish if best matching with respect to dilatations is applied. It is especially interesting that vanishing of the energy is simultaneously enforced. The reason for this is the drastic consequence of scale invariance. The point is that the kinetic energy has dimensions of length squared, and even under a $\lambda$-independent dilatation $\textbf x_i\rightarrow D\textbf x_i$ changes by a factor $D^2$. This already means that the structure of the Jacobi metric, with kinetic metric that describes pure inertial motion and a conformally related metric that describes inertial motion modified by forces, is not possible on shape space Q$_{\scriptsize\textrm{SS}}$ (though it is still possible on Q$_{\scriptsize\textrm{RCS}}$). One cannot construct a metric on shape space without a compensating potential term $F$ that is homogeneous of degree -2 and therefore transforms as $D^{-2}$, thereby compensating the $D^2$ of the kinetic term. This means that, in contrast to the Jacobi action, one cannot have a conformal factor of the form $E-U$ made up of the constant total energy $E$ and a `proper' potential that depends on the inter-particle separations.

\section{Hidden Scale Invariance}

If we are to take scale invariance seriously, we must now confront the problem that the standard potentials in Newtonian dynamics, for gravity and electrostatics, derive from potentials homogenous of degree -1, not -2. Nature would appear to be sending us a strong signal that it is not scale invariant. There is, however, a possibility that scale invariance is realized but hidden remarkably effectively. We have seen above in (\ref j) that the time derivative of the moment of inertia (\ref{MOI}) is the expansive momentum $v$. Therefore, if $v$ vanishes the moment of inertia becomes a conserved quantity. Within standard Newtonian theory, this is an exceptional case, requiring the simultaneous vanishing of the energy and the expansive momentum. It is, however, a necessary consequence of scale invariance as defined here. Let us therefore exploit this fact by converting given Newtonian potentials into scale-invariant analogues that have the necessary homogeneity of degree -2. To do this, we shall use $I$, or rather $\sqrt {MI}$:
\begin{equation}
    \mu=\sqrt{\sum_{i<j}m_{i}m_{j}r_{ij}^{2}}.
\label{Rho}
\end{equation}

Just as one passes from special to general relativity (with gravity minimally coupled to matter) by replacing
ordinary derivatives in the matter Lagrangians by covariant
derivatives, Newtonian potentials can be converted into potentials
that respect scale invariance. One simply multiplies by an
appropriate power of $\mu$, which has the dimensions of length.
This is a rather obvious mechanism. What is perhaps unexpected is
that the modified potentials lead to forces $\textit{identical}$
to the originals accompanied by a universal cosmological force
with minute local effects. The scale invariance is hidden because
$\mu$ is conserved.

Let some standard Newtonian potential $U$ consist of a sum of potentials $U_{k}$ each homogeneous of
degree $k$:
\begin{equation}
    U=\sum_{k=-\infty}^{\infty}a_{k}U_{k}.
\label{NewtPot}
\end{equation}

The $a_{k}$ are freely disposable strength constants. The energy
$E$ in the Jacobi action (\ref{Jacobi}) will be treated as a
constant potential ($k=0$). (It plays a role like the cosmological
constant $\Lambda$ in GR).

Now replace (\ref{NewtPot}) by
\begin{equation}
    \tilde{U}=\sum_{k=-\infty}^{\infty}b_{k}U_{k}\mu^{-(2+k)}.
\label{ScaledPot}
\end{equation}

The equations of motion for (\ref{NewtPot}) are
$$
    {\textrm{d}\textbf{p}^{i}\over\textrm{d}t}=
    -\sum_{k=-\infty}^{\infty}a_{k}{\partial
    U_{k}\over\partial\textbf{x}^{i}};
$$
for (\ref{ScaledPot}) they are
$$
    {\textrm{d}\textbf{p}^{i}\over\textrm{d}t}=
    -\sum_{k=-\infty}^{\infty}b_{k}\mu^{-(2+k)}
    {\partial U_{k}\over\partial\textbf{x}^{i}}+
    \sum_{k=-\infty}^{\infty}(2+k)b_{k}\mu^{-(2+k)}U_{k}
    {1\over\mu}{\partial\mu\over\partial\textbf{x}^{i}}.
$$

Since $\mu$ is constant `on shell', we can define new strength
constants that are truly constant:
\begin{equation}
    b_{k}=a_{k}\mu^{2+k}.
\label{DefB}
\end{equation}
The equations for the modified potential become
\begin{equation}
    {\textrm{d}\textbf{p}^{i}\over\textrm{d}t}=
    -\sum_{k=-\infty}^{\infty}a_{k}
    {\partial U_{k}\over\partial\textbf{x}^{i}}+
    \sum_{k=-\infty}^{\infty}(2+k)a_{k}U_{k}
    {1\over\mu}{\partial\mu\over\partial\textbf{x}^{i}}.
\label{ModEq}
\end{equation}
If we define
\begin{equation}
    C(t)={\sum_{k=-\infty}^{\infty}(2+k)a_{k}U_{k}\over
    2\sum_{i<j}m_{i}m_{j}r_{ij}^{2}}
\label{DefC}
\end{equation}
and express $\mu$ in terms of $r_{ij}$, then equations
(\ref{ModEq}) become
\begin{equation}
    {\textrm{d}\textbf{p}^{i}\over\textrm{d}t}=
    -\sum_{k=-\infty}^{\infty}a_{k}
    {\partial U_{k}\over\partial\textbf{x}^{i}}+
    C(t)\sum_{j}m_{i}m_{j}{\partial{r_{ij}^{2}}\over\partial\textbf{x}^{i}}.
\label{AbbModEq}
\end{equation}

We recover the original forces exactly together with a universal
force. It has an epoch-dependent strength constant $C(t)$ and
gives rise to forces between all pairs of particles that, like
gravitational forces, are proportional to the inertial mass but
increase in strength linearly with the distance.\footnote{Although $C(t)$ is epoch dependent, this does not mean
that the theory contains any fundamental coupling constants with such a dependence. The epoch dependence is
an artefact of the decomposition of the forces into Newtonian-type forces and a residue, which is the
cosmological force.} The universal
force will be attractive or repulsive depending on the sign of
$C(t)$, which is an explicit function of the $r_{ij}$'s. For small
enough $r_{ij}$, the force will be negligible compared with Newtonian
gravity. However, on cosmological scales it will be significant. I refer the reader to \cite{SIG} for a discussion of the possible cosmological implications. Since this paper is primarily concerned with an alternative formulation of the relativity principle and the problem of time, let me conclude this section with some comments about these two issues and then, in the next section, describe what happens in the context of Riemannian geometry and field theory.

The main significance of the model presented here is, I believe, methodological. It shows that one can formulate a powerful constructive relativity principle and implement it universally. One postulates a spatial geometry and the nature of the objects it. Then the geodesic principle and best matching lead to a highly predictive theoretical framework that is completely relational. Above all, observable effects have genuine observable causes. This is because the kinematic structure that is presupposed is pared down to the bare minimum: spatial geometrical relationships quotiented with respect to the spatial symmetries. No extra kinematics associated with time and inertial systems is assumed. Nevertheless, Newton's laws with extra restrictions that do not appear to be in conflict with observations are recovered. The theory clearly cannot fix everything despite the strong restrictions. The potential can still contain several independent terms with arbitrary relative strengths. However, effects without observable material causes are completely eliminated. Specifically, there are no observable effects that one could attribute to translation, rotation or change of size of the complete universe. These results were to be expected, but, very interestingly, the imposition of scale invariance also eliminates the last vestige of what one might call time kinematics: the possibility of including the constant total energy $E$ in Jacobi's principle. In Newtonian theory, different values of $E$ are possible because the external absolute time allows different kinetic energies for a given spatial configuration. Absolute time does seem to be abolished from Jacobi's principle, but its effect is exactly reproduced by the freely specifiable constant $E$ that is not associated with observable sources. There is still a free constant $E$ in the scale-invariant theory, but it is now the coefficient of a genuine potential. 

Scale invariance is a highly intuitive and theoretically desirable attribute of any dynamics of the universe. I find it remarkable that it also `kills time.' It has even more striking consequences in the context of Riemannian geometry and field theory, to which we now turn.

\section{Riemannian Geometry and Fields}

This section merely serves as a summary of the content of the papers \cite{BOF, AB, ABFO, Variations} with emphasis on the connection between time and scale invariance. I shall start by reviewing the manner in which general relativity, which was, of course, originally formulated as a description of four-dimensional pseudo-Riemannian spacetime, can be interpreted as a dynamical theory of the evolution of three-dimensional Riemannian geometry.

A given Einsteinian spacetime, which for simplicity I shall assume has compact spatial sections (closed universe),  can, provided it is globally hyperbolic, be foliated by spacelike hypersurfaces (`leaves'), on which a Riemannian geometry is induced. Such foliations can be generated by laying down some system of coordinates on the original spacetime. Then the surfaces of constant value of the time coordinate are the hypersurfaces of the foliation, on which the induced Riemannian geometry is represented by a Riemannian 3-metric $g_{ij}$ defined relative to the spatial coordinates on the given leaf. Such 3-metrics constitute the dynamical variables when general relativity is treated as Hamiltonian theory. The space of all Riemannian 3-metrics $g_{ij}$ on a given 3-manifold $\cal M$ is called Riem($\cal M$). It is the analogue of the Newtonian configuration space. All 3-metrics related to each other by (spatial) coordinate transformations, or equivalently 3-diffeomorphisms, correspond to a given 3-geometry. The space of all 3-geometries (on a given manifold) is called \textit{superspace}. Mathematically, superspace is the quotient of Riem with respect to 3-diffeomorphisms. There is an obvious direct analogy between 3-diffeomorphisms, which `drag the contents of the universe' around on $\cal M$, and translations and rotations on Euclidean space. Thus, superspace is the analogue of the relative configuration space (RCS) of the particle model. 

Given the spacetime and the foliation, each induced 3-metric will be a point in Riem, and the spacetime will be a curve in Riem. Keeping the foliation unchanged (i.e., the t=constant surfaces the same), but changing the spatial coordinates freely on the different slices, one obtains many different curves in Riem that all represent the same spacetime. On superspace, for the given foliation, there is just one curve. However, if one changes the foliation, one obtains a whole family of curves in superspace, one for each foliation. This multiplicity of curves, all representing the same spacetime, reflects the relativity of time in Einstein's theory and has hitherto presented insuperable obstacles to the attempts at a canonical quantization of general relativity. This is \textit{the problem of time} \cite{Kuchar, Isham}. It is closely related to the question of the true degrees of freedom of the gravitational field. It is generally accepted that in Einstein's theory there are two at each space point. The question is: can one identify them? When general relativity is treated as a dynamical theory, the $g_{0\nu}, \nu=0,1,2,3,$ components of the spacetime metric turn out to be Lagrange multipliers and are not proper degrees of freedom. The remaining six components $g_{ij}$ in the spatial part of the metric contain three arbitrary functions, corresponding to the possibility of making arbitrary coordinate transformations. This gauge freedom is quotiented out in the passage from Riem to superspace. At this level, one is left with three degrees of freedom per space point. However, the freedom to choose the time coordinate arbitrarily, changing thereby the foliation, represents a further gauge function per space point. But what then is evolving? If one wishes to maintain full general covariance, one cannot quotient away that freedom as one does in the passage from Riem to superspace.

However, if one is prepared to sacrifice general covariance, an obvious step is to perform a further quotienting like the passage from the RCS to shape space. The analogue of shape space is \textit{conformal superspace} (CS), which is obtained by quotienting Riem not only by 3-diffeomorphisms but also by conformal transformations:
\begin{equation}
g_{ij}\rightarrow\phi^4g_{ij},
\label{k}
\end{equation}
where the fourth power of the positive function $\phi$ is chosen for mathematical convenience. 

Many years ago, York \cite{York} showed that, indeed, one can parametrize the solutions of the initial-value constraints of general relativity by the two degrees of freedom per space point that reside in conformal superspace. Furthermore, he made effective use of the conformal transformations (\ref k) to find such solutions. This important piece of work involved a \textit{constant-mean-curvature} (CMC) foliation. This is defined as follows. At each point on any leaf of a foliation, the leaf has a certain extrinsic curvature tensor $K_{ij}$ (second fundamental form). This measures the manner in which the leaf is curved in the spacetime in which it is embedded. A CMC foliation is one for which the trace of $K_{ij}$, $K=g_{ij}K^{ij}$, where indices are raised and lowered by means of $g_{ij}$ and its inverse, is constant on each leaf. For spatially compact solutions, a CMC foliation is unique, and $K$ varies monotonically with the cosmic time. CMC foliations have many useful properties and are uniquely helpful in York's method for finding solutions to the initial-value constraints of general relativity. However, because York did not arrive at his technique through a fundamental variational principle but merely exploited what has proved to be very convenient mathematics, the use of CMC foliations is regarded as a gauge-fixing condition that breaks four-dimensional general covariance. It `fixes time', introducing a definition of simultaneity, which is anathema to many relativists. However, the 3-space approach suggests an interesting alternative interpretation in which the CMC foliation has a deep physical significance. 

The stimulus to the development of the 3-space approach was the Lagrangian reformulation of the Dirac--ADM Hamiltonian representation of general relativity \cite{DiracHam, ADM} found by Baierlein, Sharp and Wheeler \cite{BSW}.
The key concepts in the ADM formalism are the 3-metrics $g_{ij}$,
the lapse $N$ and the shift $N^{i}$. The lapse measures the rate
of change of proper time w.r.t. the label time, while the shift
determines how the coordinates are laid down on the successive
3-geometries. Prior to the transition to the Hamiltonian, the
standard Hilbert--Einstein action for matter-free GR is rewritten,
after divergence terms have been omitted, in the 3+1 form
\begin{equation} I= \int \textrm dt\int\sqrt{g}N\left[R + K^{ij}K_{ij} - K^2\right]\textrm d^3x. \label{first}
\end{equation}

Here $R$ is the three-dimensional scalar curvature, and $K_{ij} =
-(1/2N)(\partial g_{ij}/
\partial t - N_{i;j} - N_{j;i})$ is the extrinsic curvature with trace $K$. From here
 the transition made by BSW
\cite{BSW} is trivial. They first replaced $K_{ij}$ in the action
by $k_{ij} = \partial g_{ij}/
\partial t - N_{i;j} - N_{j;i}$, the unnormalised normal
derivative, to give
\begin{equation}
I= \int \textrm dt\int\sqrt{g}\left[NR + {1 \over 4N}\left(k^{ij}k_{ij} - k^2\right)\right]\textrm d^3x. \label{second}
\end{equation}
They varied this action with respect to the lapse and found an
algebraic expression for it,
 \begin{equation} N =
\sqrt{{k^{ij}k_{ij} - k^2 \over 4R}}. \label{third}
\end{equation}
This expression for $N$ is substituted back
into Eq.(\ref{first}) to obtain the BSW Lagrangian
\begin{equation} 
I_{\scriptsize\textrm{BSW}} = \int\textrm dt\int {\cal L}\textrm d^3x= \int\textrm dt\int\sqrt{g}\sqrt{R}\sqrt{k^{ij}k_{ij} -
k^2}\textrm d^3x. \label{fourth} \end{equation}

This action is closely analogous to the Jacobi action (\ref{Jacobi}): it has a geodesic-type square root and the Lagrange multiplier $N_i$ generates 3-diffeomorphisms in the same way that the multipliers in the particle model generate the Euclidean symmetry transformations. These properties lead, respectively, to the quadratic and linear constraints
\begin{equation}
    -p^{ij}p_{ij} + {1 \over 2}p^2  + gR = 0,\hspace{.5cm}p^{ij}_{~~;i} = 0,\hspace{.5cm}p = g_{ab}p^{ab}, \label{2.3}
\end{equation}
where the canonical momenta are
\begin{equation}
    p^{ij} = {\delta {\cal L} \over \delta\left({\partial g_{ij} \over
\partial \lambda}\right)}, \label{2.2}
\end{equation}
and $p=g_{ij}p^{ij}$ is their trace. It measures the expansion of space and is the analogue of the expansive momentum in the particle model.

The constraints (\ref{2.3}), which are the fundamental ADM Hamiltonian and momentum constraints, are evidently like the particle constraints but with a crucial difference: instead of being global constraints that hold for the complete universe, these are infinitely many constraints, one per space point. Note, in particular, that the quadratic constraint is an identity that follows from the mere form of the Lagrangian. It is important that the square root is `local', i.e., it is taken before the integration over space. This ensures that there is a constraint per space point.

In \cite{BOF, AB, SIG, ABFO}, my collaborators and I studied systematically gravity--matter-field Lagrangians that are natural generalizations of the BSW Lagrangian and have the form
\begin{equation} 
I = \int\textrm dt\int\sqrt{g}\sqrt{U_{\scriptsize\textrm g}+U_{\scriptsize\textrm m}}\sqrt{T_{\scriptsize\textrm g}+T_{\scriptsize\textrm m}}\textrm d^3x. 
\label{l} 
\end{equation}
Here, the gravitational potential term $U_{\scriptsize\textrm g}$ depends only on the 3-metric $g_{ij}$, while the matter potential terms $U_{\scriptsize\textrm m}$ for the considered scalar and 3-vector fields depend on $g_{ij}$, the matter fields and their spatial derivatives. The kinetic terms are quadratic in the velocities, as in conventional Hamiltonian field theory, but with the all-important difference that the velocities are `corrected' to take into account the effect of time-dependent diffeomorphisms and conformal transformations. The corrections are uniquely determined by the symmetry transformation and introduce corresponding Lagrange multipliers and linear momentum constraints, just as in the particle model. However, we included a free coefficient $A$ in $T_{\scriptsize\textrm g}=k^{ij}k_{ij} -Ak^2$ to reflect that, a priori, two independent scalars can contribute to the kinetic term. I will first discuss our results for the matter-free case and best matching with respect to diffeomorphisms.

They revealed an important difference from the particle case, in which propagation of the constraint linear in the momentum imposes conditions on the potential but the quadratic constraint propagates with any potential. In the case of the Riemannian symmetry and the local square root of the BSW action, the two constraints are `intertwined' and the simultaneous propagation of them imposes strong restrictions. We found that the only consistent Lagrangians of the form (\ref l) must have the free coefficient $A$ equal to unity, as in general relativity, while the gravitational potential term must have the form $U_{\scriptsize\textrm g}=\Lambda+sR, s=0, 1, -1$. The freely specifiable constant $\Lambda$ corresponds to the energy $E$ in Jacobi's principle and is Einstein's cosmological constant. The three possible values of $s$ correspond respectively to so-called strong gravity and general relativity with spacetime signatures -+++ and ++++.  This meant that we had found a completely new derivation of general relativity by consistent application of the timeless generalization of Poincar\'e's relativity principle applied in the case of Riemannian 3-geometries. Spacetime was in no way presupposed. It was derived.

Even more remarkable results came when we tried to couple scalar and 3-vector fields to gravity in the case of the Minkowskian signature -+++. In the standard spacetime approach, the fact that such fields must respect the same light cone as the gravitational field is put in the form of the assumption that spacetime in the small is Minkowskian. A universal light cone is presupposed. However, we found that it is enforced by consistent propagation of the two constraints, both of which are modified by the addition of matter terms. A key point is that the form of the momentum constraint is always completely determined by the tensorial nature of the fields and the fact that one is best matching with respect to diffeomorphisms, which affect all fields in a uniquely determined way. In contrast, the form of the quadratic constraint reflects the particular ansatz made for the Lagrangian. Only very special Lagrangians lead to consistent constraint propagation. Thus, our first result for matter was that there must be a universal light cone. Equally striking is the fact that in the case of 3-vector fields the very same requirement of constraint propagation forces the 3-vector fields to be gauge fields. In fact, the universal light cone and gauge theory are shown to have essentially the same origin. Of course, all derivations in theoretical physics include simplicity assumptions, either explicitly or implicitly. These are discussed by Anderson in \cite{Variations}. I think it is correct to say that the 3-space approach assumes less and derives more than the standard approach based on Einstein's general principle of covariance.

So far, I have described only the effect of the local square root and best matching with respect to diffeomorphisms. In the more recent paper \cite{ABFO}, we constructed Lagrangians in which best matching with respect to conformal transformations is also performed. We obtained two main results. First, if one best matches the BSW Lagrangian as it stands with respect to conformal transformations that preserve the spatial volume $V=\int\sqrt g\textrm d^3x$, then the standard Dirac--ADM constraints are augmented by the constraint
\begin{equation}
{p\over\sqrt g}=constant,
\label{m}
\end{equation}
and one also obtains a condition on the lapse that ensures the propagation of this constraint by the Euler--Lagrange equations. When expressed in terms of the extrinsic curvature $K$, the constraint (\ref m) is precisely York's CMC slicing condition. (Note that $p/\sqrt g$ is constant on each leaf, but its value changes under the evolution.) We therefore have the striking result that such conformal transformations, which change all local scales completely freely subject to the single global restriction on the volume, lead us to general relativity in a distinguished foliation, i.e., to a distinguished definition of simultaneity. Once again, we find a strong connection between time and scale invariance. I believe that this result should be taken seriously. In general relativity, there are four gauge freedoms. Three are associated with 3-diffeomorphisms and the fourth with arbitrary transformations of the time coordinate. In our best-matching approach, there are also four gauge freedoms. The 3-diffeomorphisms are still present, but the freedom in the time gauge is replaced by freedom in the scale gauge. All the four gauge freedoms are now expressed through constraints linear in the canonical momenta. Moreover, all have their origin in the geometry of space.

However, from the point of view of scale invariance, general relativity is frustratingly not quite perfect. I find it extremely puzzling that the solitary volume-preserving restriction on full scale invariance is imposed on the conformal transformations. In fact, this is what permits volume to be a physical degree of freedom in general relativity and allows the expansion of the universe. I already mentioned that the trace $p$ of the canonical momenta measures the expansion of space. We see from the constraint (\ref m) that $p$ does not vanish but that $p/\sqrt g$ is equal to an evolving spatial constant. When we best match with respect to all conformal transformations, dropping the volume-preserving restriction, we get
\begin{equation}
p=0.
\label{n}
\end{equation}

There is now full agreement with the scale-invariant particle model, for which the vanishing of the expansive momentum ensures constancy of the moment of inertia (the `size' of the $N$-particle universe). Here, the vanishing of $p$ means that the volume of the universe cannot change. Because the BSW Lagrangian (\ref{fourth}) allows the volume to change, it has to be modified if the constraint $p=0$ is to propagate. Just as constancy of the moment of inertia was achieved by dividing the potential by a power of the moment of inertia, we achieved propagation of $p=0$ by dividing the gravitational potential by an appropriate power of the volume. This led us to a consistent fully scale-invariant theory that we call \textit{conformal gravity}. It is a remarkably small modification of general relativity. The single global variable that permits the volume of the universe to change is excised and one obtains a dynamics of the geometry of pure shape. Unfortunately, although conformal gravity should describe the solar-system and binary-pulsar data just as well as general relativity, the cosmology must be quite different. At the time of writing, it seems hard to believe that conformal gravity will be able to supplant general relativity as a cosmological theory. The difficulties are spelled out in \cite{SIG, ABFO}.

However, this probable failure of conformal gravity does not change the fact that the basic idea of using best-matching geodesics to implement Mach's principle on the basis of Poincar\'e's relativity principle establishes unexpected links between special relativity, general relativity and the gauge principle. They all emerge together as a self-consistent complex from a unified and completely relational approach to dynamics. We see that all of currently dynamics can be understood in terms of purely spatial geometry. Finally, a deep connection between time and shape is established.

It was a great pleasure to participate in DICE2002, and I hope this contribution to the proceedings will foster further interdisciplinary workshops.


\begin{thebibliography}{99}
\scriptsize

\bibitem{Leibniz}   Alexander H G (ed.) 1956 \textit{The Leibniz--Clarke Correspondence}, Sec. 47 (Barnes and Noble, New                 York)
\bibitem{Mach}      Mach E 1883 \textit{Die Mechanik in ihrer Entwicklung historisch-kritsch
                    dargestellt} (Leipzig: Barth); 1893 \textit{The Science of
                    Mechanics} (Chicago: Open Court)
\bibitem{Poincare1} Poincar$\acute{\textrm e}$ H 1905 \textit{Science and
                    Hypothesis} (London, translated from the French edition of 1902)
\bibitem{AutoNotes} Einstein A 1949 ``Autobiographical Notes,'' in \textit{Albert Einstein -- Philosopher -- Scientist}, ed. P A Schilpp (The Library of Living Philosophers: Evanston, Illinois)
\bibitem{Barbour}   Barbour J 1990 ``The part played by Mach's Principle in the genesis of relativistic cosmology'' in \textit{Modern Cosmology in Retrospect}, eds. B. Bertotti, R. Balbinot, S. Bergia, and A. Messina (Cambridge University Press: Cambridge) (see also: Barbour J 1999 ``The development of Machian themes in the twentieth century'' in \textit{The Arguments of Time}, ed. J Butterfield (Oxford University Press: Oxford)).
\bibitem{Einstein16}  Einstein A 1916 \textit{Annalen der Physik} \textbf{49}, 769
\bibitem{Kretschmann}   Kretschmann E 1917 \textit{Annalen der Physik} \textbf{53}, 575 
\bibitem{Einstein18}  Einstein A 1918 \textit{Annalen der Physik} \textbf{55}, 241
\bibitem{BarPlanck} Barbour J 2001 ``On general covariance and best matching'' in \textit{Physics Meets
                    Philosophy at the Planck Length}, eds. Callender C and Huggett N (Cambridge: Cambridge University
                    Press)
\bibitem{BOF}       Barbour J, Foster B Z and \'{O}
                    Murchadha N 2002 \textit{Class. Quantum Grav.} \textbf{19} 3217;
                    ``Relativity without relativity'', arXiv:gr-qc/0012089
\bibitem{AB}        Anderson E and Barbour J 2002 \textit{Class. Quantum Grav.}
                    \textbf{19} 3249; ``Interacting vector fields
                    in relativity without relativity'',
                    arXiv:gr-qc/0201092
\bibitem{SIG}       Barbour J 2002 ``Scale-invariant gravity: particle dynamics'', arXiv:gr-qc/0211021 (to be published in \textit{Class. Quantum Grav.}

\bibitem{ABFO}      Anderson E, Barbour J,
                    Foster B Z and \'{O} Murchadha N 2002 ``Scale-Invariant Gravity:
                     Geometrodynamics", arXiv:gr-qc/0211022 (to be published in
                    \textit{Class. Quantum Grav.}
\bibitem{Lanczos}   Lanczos C 1949 \textit{The Variational Principles of Mechanics} (University of Toronto Press,
                    Toronto)(also available from Dover, New York)
\bibitem{BB}        Barbour J and Bertotti B 1982 \textit{Proc. R. Soc. A} \textbf{382} 295    
\bibitem{CQG94}     Barbour J 1994 \textit{Class. Quantum Grav.} \textbf{11} 2875
\bibitem{EOT}       Barbour J 1999 \textit{The End of Time} (London: Weidenfeld and Nicolson; New York: Oxford University
                    Press)
\bibitem{Dirac}     Dirac P A M 1964 \textit{Lectures on Quantum Mechanics} (New York: Yeshiva University)
\bibitem{Poincare2} Poincar\'e H 1898 \textit{Rev. M\'etaphys. Morale} \textbf{6} 1 (English translation 1913: 
                    ``The Measure of Time'' in \textit{The Value of Science} (New York: Science Press)
\bibitem{Clemence}  Clemence G 1957 \textit{Rev. Mod. Phys.} \textbf{29} 2
\bibitem{Variations} Anderson E 2003 ``Variations on the seventh route to relativity,'' arXiv:gr-qc/0302035
\bibitem{Kuchar}    Kucha\v r K 1992 ``Time and interpretations of quantum gravity'' in \textit{Proceedings of the 4th Canadian Conference on General Relativity and Relativistic Astrophysics} eds. G Kunstatter, D Vincent and J Williams (World Scientific: Singapore) 
\bibitem{Isham}     Isham C 1992 ``Canonical quantum gravity and the problem of time'' in \textit{Integrable Systems, Quantum Groups, and Quantum Field Theories} eds. L Ibart and M Rodriguez (Kluwer: Amsterdam)
\bibitem{York}      York J 1971 \textit{Phys. Rev. Letters} {\bf 26}, 1656; York J 1972 \textit{Phys. Rev. Letters}
{\bf 28}, 1082
\bibitem{DiracHam}  Dirac P 1958 \textit{Proc. Roy. Soc. Lond.} {\bf A246}, 333

\bibitem{ADM}       Arnowitt R, Deser S, Misner C 1972 ``The dynamics of general relativity'' in \textit{Gravitation: an
Introduction to Current Research}, ed. L Witten (Wiley: New York)
\bibitem{BSW}       Baierlein R, Sharp D and Wheeler J 1962 \textit{Phys. Rev.} {\bf 126},
1864



\normalsize


\end{thebibliography}
\end{document}